# A Software Radio Challenge Accelerating Education and Innovation in Wireless Communications

Marta González-Rodríguez, Antoni Gelonch-Bosch
Dept. Signal Theory and Communications
Barcelona Tech-UPC, Castelldefels, Spain
{marta.gonzalez.rodriguez|antoni}@tsc.upc.edu

Vuk Marojevic
Dept. Electrical and Computer Engineering
Mississippi State University, Mississippi State, MS
vuk.marojevic@gmail.com

*Abstract*—**This Innovative Practice Full Paper presents our methodology and tools for introducing competition in the electrical engineering curriculum to accelerate education and innovation in wireless communications. Software radio or software-defined radio (SDR) enables wireless technology, systems and standards education where the student acts as the radio developer or engineer. This is still a huge endeavor because of the complexity of current wireless systems and the diverse student backgrounds. We suggest creating a competition among student teams to potentiate creativity while leveraging the SDR development methodology and open-source tools to facilitate cooperation. The proposed student challenge follows the European UEFA Champions League format, which includes a qualification phase followed by the elimination round or playoffs. The students are tasked to build an SDR transmitter and receiver following the guidelines of the long-term evolution standard. The metric is system performance. After completing this course, the students will be able to (1) analyze alternative radio design options and argue about their benefits and drawbacks and (2) contribute to the evolution of wireless standards. We discuss our experiences and lessons learned with particular focus on the suitability of the proposed teaching and evaluation methodology and conclude that competition in the electrical engineering classroom can spur innovation.**

## I. INTRODUCTION

Wireless technologies and procedures, as well as design and deployment aspects have been present in the electrical engineering curriculum for decades. When software radios or software-defined radios (SDRs) became feasible, widely available and affordable, instructors started using software toolboxes and SDR hardware peripherals to teach wireless communications principles, standards, and practical implications. With SDR available tools, students can take the role of the radio system or component developer and use well-established benchmarks to engineer their system.

Active learning (AL) methodologies, where students get conscious and take responsibility about their own learning progress, were introduced in university courses. These methodologies can address different working principles: cooperative, competitive, individual, objective-based or grading centric learning. AL methodologies exhibits better students comprehension levels, encourages deep learning and help students to acquire and retain information [1]–[4]. When properly implemented, AL improves inter-group cooperation and intra-group coordination, management of multidisciplinary teams, leadership, achievement of set requirements and deadlines and self-assessment of accomplished work; all these are important skills for future industry employees [4]–[6].

Project-based learning (PBL), where students learn while developing a project, is a well know cooperative learning technique that can naturally leverage SDR technology. But, PBL is not the panacea. We have used PBL methodologies in our classes for more than 10 years [7] and on several occasions we observed a lack of motivation and participation. Concretely, some students or teams do well while others simply follow, or give up. More precisely, only few students really work, progress and learn while others observe. This is disappointing for the instructor and unfair to the active students. Other examples, both positive and negative, exist.

As stated by [8], an effective classroom should have the right mix of cooperative, but also competitive and individualized learning. Competitive learning, where students compete for some well-established objective, not only increase motivation, but also improves creativity and strengthens collaboration capabilities. Moreover, most students will collaborate, but also compete at different levels in their professional career: with their colleagues in the company for personal success and with other companies to position their products in the market. With all that said, it seems necessary for universities to find the right balance between the benefits and drawbacks of competitive learning and introduce real-world industry or research challenges in the engineering curriculum.

We propose a competitive PBL approach, a new hands-on methodology for teaching cellular communications. We adopt competitive learning and develop a challenge following the format of the European UEFA Champions League. This championship includes a qualification phase that is followed by the playoffs. Students, in small teams develop a simplified $4^{th}$ generation (4G) long-term evolution (LTE) radio link in software. They are given a framework, which provides the general code structure with partially implemented functions. Using this template, each team develops its personalized LTE system and optimizes it to outperform other teams' systems in terms of metrics that are defined at the beginning of class.

We leverage SDR technology to develop the challenge. SDR enables rapid and low-cost prototyping, testing and deployment, which are very interesting features for innovation in the classroom. Many free open-source software tools and af-fordable hardware components emerged in the last decade. One popular software framework is GNU Radio, which is used in research and education. GNU Radio facilitates interfacing with popular SDR hardware, such as the Universal Software Radio Peripherals (USRPs). Other software libraries that facilitate building radios include liquidDSP and srsLTE, both developed in C for optimized implementation and supporting real-time execution of complex wireless systems, such as generic OFDM waveforms and 3rd Generation Partnership Project (3GPP) compliant LTE systems.

This paper presents our approach to competitive learning in combination with cooperative learning under the PBL umbrella when applied to *Software Radio Engineering (SRE)*, which is a compulsory third year, second semester B.Sc. course in Electrical Engineering taught at the Telecommunication and Aerospace Engineering School of Castelldefels (EETAC), Barcelona Tech-UPC. A successful intervention requires the careful definition of the learning objectives, teaching methodology and evaluation process. We discuss these aspects and analyze the validity and effectiveness of our approach. The rest of the paper is organized as follows: Section II reviews the state of the art of related learning methods. Section III describes the learning objectives and teaching methodology and the challenge organization. Section IV presents the SDR tools that we developed and used for this intervention. Section V analyzes the suitability of using a combined cooperative-competitive methodology from different angles. Section VI provides further discussion and the conclusions.

## II. LITERATURE REVIEW

### A. Cooperative Learning

Cooperation is defined as the act of working together to accomplish common goals. Cooperative learning is the pedagogical use of this concept, where small groups of students work together to maximize their own learning success and that of others. Students therefore share information, activities and efforts to reach the proposed objective. It has been a widely used teaching strategy in the recent past [8][9] and some of the existing literature argues that cooperative schemes strongly increase the student motivation and, therefore, the learning outcomes. The principle of cooperative learning is the creation of a classroom atmosphere that invites active participation in course activities. This is especially helpful for students who struggle with the subject as they will receive the guidance from more experienced students. Nevertheless, cooperative learning can fail simply because of the personality of participants and their different backgrounds. Moreover, leadership may be taken by few students who dominate discussions and take decisions while others do not engage at all. The probably most significant drawback of cooperative learning is the imbalanced contribution to the team's goals, resulting from different engagements. In other words, as opposed to leveraging the diverse and cooperative learning environment, it is often the case that only a few students progress and achieve the learning objectives.

### B. Project-Based Learning

Learning by doing has been a major breakthrough in engineering education for the past decades. It is inspired by how humans learn, how they develop expertise and what mechanisms they activate when thinking at higher level [6]. Students develop their projects and, as part of the process, acquire valuable technical and nontechnical skills and explore the subject in more detail. These projects are often defined around simplified real-world problems.

Students usually work in groups where each individual adds value to the project. A diverse group is rich in different experiences, abilities, learning styles and perspectives. Depending on the circumstance, the instructor may define the conceptual problem and how to address it. The true value of PBL comes from the students being treated as professionals and let them investigate like in a professional R&D environment.

PBL also has its drawbacks, which were observed when students lack sufficient background knowledge [10]. In such cases, they experience difficulties to initiate their project and usually do not reach the necessary depth to succeed. To overcome this, [11] proposes the spiral step-by-step method, where information is grouped into stages and provided sequentially so that students can better focus and develop the necessary background with enough depth. Past experiences have shown that PBL benefits can be found in robotics [12][13], embedded systems [14], and computer architecture [15] education as well as in multidisciplinary subjects, such as electronics and instrumentation [16] or programming, embedded systems and introductory robotics [17].

### C. Competitive Learning

Competition is an activity of two or more parties that are attempting to achieve an exclusive goal, which cannot be held in common or shared among the parties. Competition usually compares performance among participants when doing the same task. It is often associated with gaming and has received significant attention in recent years in education [18][19]. There are positive, but also negative aspects of competitive learning. While some authors argue that competition does not increase motivation [20] and do not recommend its use [21][22], others argue that it does [23][24]. The opposed opinions have come closer over the years and competition is accepted today to complement other learning approaches if its negative effects can be minimized. The negative effects include disappointments because of lack of success and the resulting lack of interest [25], reduction of self-confidence, poor attitude towards errors [26][27], and loss of efficacy [28]. Nevertheless, in literature there exist a plethora of competitive learning studies showing its benefits [21][25][29][30][31]. Others address how to mitigate the negative effects. In particular, [32] compares different competition approaches like face-to-face, decreased proximity (not face-to-face, but with competitor's identity known), and anonymous. In [33], the "learning by losing" experience is put in value as it prepares students for working under pressure. Reference [34] compares two gaming approaches, competitive and cooperative. The authors do not find significant differences in the learning effectiveness and learning motivation, although competition seems to work slightly better than cooperation.

*D. Cooperation Plus Competition*

Studies exist that combine collaboration and competition. Relevant to our work is [35], which describes a methodology where students work in project teams and participate in a competition that is similar to our Champions League approach, with qualification and elimination rounds. Their results show that combining competitive and cooperative learning has the potential to become a solid and reliable engineering education tool, inspiring our research.

## III. LEARNING OBJECTIVES AND TEACHING METHODOLOGY

The SRE course is offered at EETAC school. It is devoted to educating students on practical aspects of modern wireless systems implementation and testing using SDR technologies and tools. Teaching state-of-the-art wireless communications systems beyond the fundamental principles is not typical in a university curriculum. But, because of the importance and growth of the wireless industry, we argue that providing a solid background for developing and managing communications systems should be part of the electrical engineering curriculum. The complexity of the latest generations of wireless systems, the diverse student population and the need to accommodate the learning process to the semester schedule makes this a challenging endeavour. In addition, new engineers must extend their skillset and add the capability to cooperation with other team members and the ability to find creative solutions that can provide a competitive advantage of the product or service being developed.

We propose four mechanisms to success: (1) create a competition among student teams to potentiate students' **motivation, creativity and effectiveness**, (2) propose a PBL methodology and use open-source tools to promote and facilitate **collaboration** and information sharing among students, (3) leverage a **system development methodology** supported by SDR tools that follows a step-by-step approach and (4) promote **engagement** by having the students actively participate in the definition of goals, etc., and involving them in the learning process.

With all these concepts in mind, the main SRE learning objectives focus on the development and management of real SDR systems with emphasis on:

a) Analyzing the impact of different building blocks, including analog-to-digital converters (ADCs), on the system performance.

b) Comparing the different amplitude, modulation and coding options offered by the LTE specifications.

c) Analyzing and comparing options for receiver synchronization and digital up and down conversion (DUC/DDC).

The class is organized around 6 theory themes and five grading activities. Along with those come two theory exams, EX1 and EX2, and 3 laboratory (lab) evaluation sessions, LAB1, LAB2, and LAB3, counting as 20%, 25%, 10%, 10% and 35% towards the final grade.

*A. Cooperative Learning Activities*

The proposed cooperative-competitive learning approach is built around labs 2 and 3, which aim at implementing a simplified version of the LTE physical layer, downlink only, and evaluate its performance in real conditions. To facilitate working on the project beyond the laboratory sessions, we substituted the radio frequency (RF) channel (requiring RF front-ends) by an audio channel, using audio devices that are available in all PCs. As a result, the project proposes using the 1.4 MHz LTE mode for reasons of low computational complexity while accommodating all system parameters to the audio sampling frequency of 48 kHz.

To accommodate the system complexity to the students' backgrounds and semester schedule, the teams are provided with an SDR framework that provides the simplified software implementation of the LTE downlink processing chain (Fig. 1) with completely (red circles) and partially (green circles) implemented modules. In order to further reduce system complexity, we implement the Primary Synchronization Signal (PSS) and LTE's physical downlink shared channel (PDSCH) only, which uses the Resource Blocks (RB) in the frame for transmitting user data.

The system performance evaluation is done using the bit error rate (BER) and block error rate (BLER) metrics. The modules **data_source_sink** and **UNCRC** compute the BER and BLER respectively. The **CHANNEL** module encompasses the **noise_channel** module (AWGN simulator) and the **DACK_JACK0** module (JACK audio driver). Moving from real implementation to simulation only requires replacing the **CHANNEL** module in the application description file.

*B. Competitive Learning Activities*

The competitive learning activities are based on the establishment of rules and a transparent grading process. We follow the European UEFA Champions League format. To pass the qualification phase, the teams need to make the LTE downlink work and support a minimum performance. While doing so the students get familiar with the LTE standard specifications. The challenge for the playoffs is to develop an optimized system that outperforms the other teams' systems in terms of specific metrics that are derived from the LTE specifications, BER and BLER, in this case. In order to be competitive, each team needs to develop its system optimization strategy which requires in a deeper understanding of the system, its parameters and dependencies.

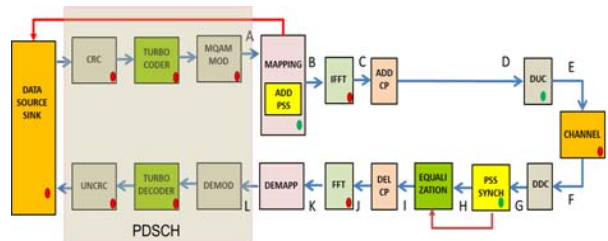

Fig. 1. Proposed LTE physical layer processing chain.

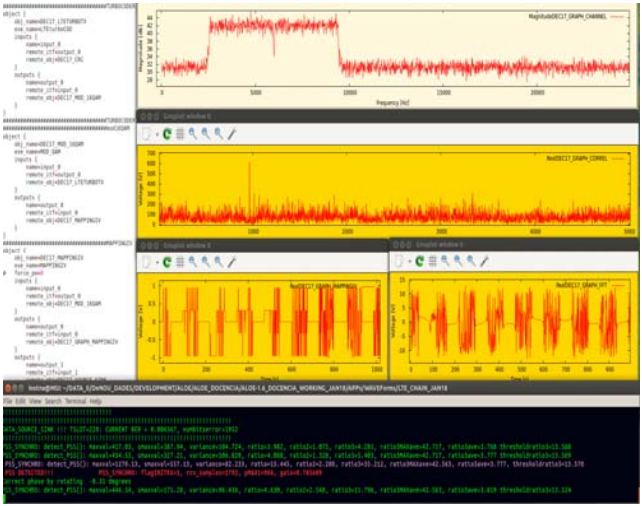

Fig. 2. ALOE working environment. The Upper-left window shows the .app file, which defines the modules and interface among them. The lower window provides execution control and system status information. The graphs show a LTE spectrum, the PSS correlation and the signals at the output of the IFFT (transmitter) and the FFT modules (receiver).

## IV. TOOLS

The abstraction layer and operating environment (ALOE) is an open-source SDR framework that is specifically designed for the implementation of modern radio systems [36]. It provides a set of customizable services for researchers, students and developers to develop, manage and evaluate waveforms in real and simulated channels. The framework abstracts heterogeneous multiprocessor platforms, provides a packet-oriented network with FIFO-based interfaces between processors, coordinates the real-time execution of the entire system and facilitates real-time monitoring of system parameters. Fig. 2 provides a screenshot of the working environment of ALOE. The processing chain, or waveform, is composed of several modules that are linked by ALOE. The waveform developer simply connects the input and output ports of the modules in the .app file, which is illustrated in the upper left part of Fig. 2. The terminal on the bottom provides execution control information, whereas the graphs visualize signals in user-defined formats. ALOE allows interfacing with external SDR peripherals, such as USRPs. Components can be easily added to interface with custom APIs to access RF transmission or acquisition boards, sound cards, and so forth.

## V. PROCEDURES

The challenge has been organized around two phases: a) Qualification and b) Playoffs. The qualifying round is embedded in LAB2 and the final round in LAB3. LAB1 is done beforehand to get familiar with the use of the ALOE development environment.

### A. Qualification

Some minimum system expectations are defined for students to fulfill the qualification. More precisely, we define BER target of 0.1 or below when operating the link with a simulated channel with a signal-to-noise ratio (SNR) of 10 dB. Passing this qualification, the students can access the playoffs. The procedure is that each team has fifteen minutes to achieve the goal on the day of the Champions event. The qualifying round grades the students according to Table I.

All the tools are provided for the groups to be successful in reaching the BER < 0.1 target with reasonable effort and, so far, all teams have been able to achieve this.

### B. Playoffs

The goal of this phase is to demonstrate high LTE link performance in real channels. The teams use the audio devices and the acoustic channel, but other scenarios are possible. A combination of scenario conditions and the achieved performance gives the Performance Score (PS) according to Table II. The final round takes a playoffs format and is organized around matches where in each match, the team with the better performance moves to the next playoff phase. The match duration is 20 minutes, which is strictly enforced. In case of equal PS scores, the team with a better overall system performance wins, where first the BER is considered and, if needed, the throughput. The LAB3 grades are assigned as shown in Table III.

The final grade for each team is calculated as a superposition of two components, the PS and the Champions Score (CS):

$$FinalMark = (\alpha \cdot CS + \beta \cdot PS), \quad (1)$$

where α and β are agreed in class at the beginning of the semester.

Each student gets an individual grade for LAB3 by multiplying FinalMark by GroupAutoEvaluation, where GroupAutoEvaluation is the average of the assessments of the team members.

### C. Step-by-Step Development

The teaching methodology follows the PLB approach. But, as commented, the complexities of modern communications systems require setting up the structure of the entire software system and providing some of the required functions. In addition, the instructor should define the development methodology to help the teams succeed. The decision on which modules to provide, partially or totally implement, should be based on the class focus and learning objectives. For example, we chose the Automatic Modulation and Coding (AMC), the digital up and down conversion (DUC/DDC) and the PSS synchronization objectives as the focal points for student development and optimization. We provided a fully operational LTE turbo encoder and decoder with the corresponding rate-matching, an implementation of the modulator and hard demodulator, an implementation of the DUC and a partial, though operational, implementation of the PSS synchronization. Students take it from there and can, with little effort, set up a complete an operational LTE downlink, but the performance may be low.

Table I. Team scoring table for qualifying round.

| BER | Mark |
|---|---|
| < 0.1 | 10.0 |
| between 0.1 and 0.2 | 8.0 |
| between 0.2 and 0.3 | 4.0 |
| between 0.3 and 0.4 | 1.0 |

Table II. Performance score.

| Scenario | BER | SNR | Distance | Performance Score |
|---|---|---|---|---|
| Audio | < 0.1 | | 1.0 m | 10 |
| Audio | 0.1<BER<0.2 | | 1.0 m | 9.5 |
| Audio | < 0.1 | | 0.25 m | 9 |
| Audio | 0.1<BER<0.2 | | 0.25 m | 8.5 |
| Cable | < 0.1 | | 1 m | 8 |
| Cable | 0.1<BER<0.2 | | 1 m | 7.5 |
| Simulation | < 0.1 | -1 dB | | 7 |
| Simulation | < 0.1 | 0 dB | | 6.5 |
| Simulation | < 0.1 | 1 dB | | 6 |
| Simulation | < 0.1 | 2 dB | | 5.5 |
| Simulation | < 0.1 | 3 dB | | 5 |
| Simulation | < 0.1 | 4 dB | | 4.5 |
| Simulation | < 0.1 | 5 dB | | 4 |
| Simulation | < 0.1 | 6 dB | | 3.5 |
| Simulation | < 0.1 | 7 dB | | 3 |
| Simulation | < 0.1 | 8 dB | | 2.5 |
| Simulation | < 0.1 | 9 dB | | 2 |
| Simulation | < 0.1 | 10 dB | | 1.5 |
| Simulation | < 0.1 | 11 dB | | 1 |
| Simulation | <0.1 | 12 dBs | | 0.5 |

We teach the students how to design, develop and verify a software-focused SDR project. We propose a detailed system development and testing methodology, where modules are tested as standalone units first before step-wise integration and testing. We also suggest the types of tests that are available and open the discussion about their pros and cons. The student teams ultimately choose the design, development and evaluation methodology that works best for them and the team goals.

As illustrated in Fig. 1 the development methodology suggests creating the application, which connects DSP modules to generate the transmitted waveform and process the received waveform. One initial way of testing is to connect the output of module A in the transmitter with the input of the mirror module, the one that does the inverse of A, at the receiver. It can be done through direct connection first, but most of the modules require a channel to assess correct and standard-compliant functioning of the transmitter-receiver pair. For example, for testing the modulator-demodulator pair, the step-by-step approach could be as follows:

Table III. Champions score.

| Championship Result | Champions Score |
|---|---|
| *Champion* | 10.0 |
| *Finalist* | 8.0 |
| *Semi-Finalist: Third, Fourth* | 7.0, 6.0 |
| *Quarter-finalist: 5th, 6th, 7th, 8th* | 5.0, 4.0, 3.0, 2.0 |

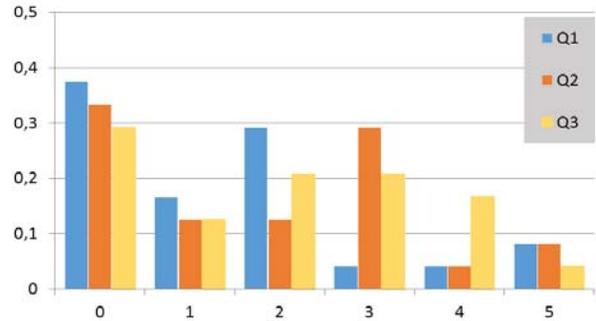

Fig. 3. Histogram of student survey results.

*Step 0. Without CRC and Turbocoder*

- Create an .app by connecting point A to point L and evaluate the BER at the output of the demodulator with an ideal channel.
- Introduce a simulated channel (noise) and verify that the BER over SNR curves are as expected.

*Step 1: With CRC and Turbocoder*

- Incorporate the CRC, UNCRC, Turbocoder and TurboDecoder and proceed as before. Notice that the coding rate needs to be taken into account.

Similar tests can be defined with other transmitter-receiver pairs. The instructor organizes monothematic sessions and provides the necessary information, covering the design and development of the LTE PDSCH (bit level processing), the Mapping and deMapping, the DUC and DDC, and the PSS signal and how to use it for synchronization. Further details are beyond the scope of this paper.

*D. Student Roles*

Increasing the student's motivation and their immersion into the learning process encouraged us to develop the competitive PBL learning methodology. In addition to the system development and testing, students need to take organizational and management roles, requiring them to take responsibility and play fair. Those activities have been organized around:

- *Rules*. Based on the instructor's initial proposal, a discussion related to the working rules and the evaluation process is initiated. Subsequent agreements are made through majority voting. The aim of these discussions is to make students more aware of the learning process and the learning methodology.
- *Event*. The challenge itself, qualification round and playoffs, takes place in a public place on campus, where anyone interested can see the different teams preparing for and competing in the matches.
- *Logistics*. Different committees are needed for running the event and ensure all the logistics are in place. The event needs speakers, someone for setting up and tearing down the infrastructure, referees, etc. A special mention needs to be made on the student referees who control the different stages of the competition and sign the acts that contain the results achieved by the teams in the different phases of the challenge.

## VI. EVALUATION

Evaluating teaching/learning methodologies is not an easy task. We need to consider the integration principles and evaluate all aspects of teaching and learning activities. The evaluation cannot rely only on experience and intuition. It should be based on scientific analysis and criteria along with scientific procedures and evaluation methods. There is a large set of methods to evaluate teaching/learning methodologies [37]. Many of them are based on student opinion as student ratings have been repeatedly shown to have a high level of validity [1]. On the other hand, it must be assumed that an instructor knows the subject, has broad background, skills and curriculum to develop appropriate teaching and learning activities and adapting them in consonance to the evolution of science and technology. Therefore, it seems reasonable to combine different approaches to evaluate an intervention or learning methodology.

### A. Surveys

Competitive learning, in use in several university programs, is still considered as an innovative technique. It is therefore important to carry out surveys on everything related to competition-based learning and accompanying evaluation. In the presented work, surveys were performed the day of the challenge, just before starting the event. We asked the students to qualify their experience in three categories with a value between 0 (worst) and 5 (best). The questions were:

Q1: Do you believe that a Champions-type challenge is adequate to learn and master the SRE subject?

Q2: Do you think that the project objectives are suitable?

Q3: Do you consider the challenge being well organized?

We received 24 answers from a total of 29 students in the course. Fig. 3 plots the distribution of answers. We observe that 29% answers are 0 (worst) to all three questions. A considerable number of students qualify the process as 2 or 3 and a few are excited about the approach. Notice that it is difficult to understand the "0 to all" answers especially when we take into consideration the one-month debate that led to the definition of details for the challenge. We noticed certain reluctance by some students to the idea of getting partial subject scores from competition with classmates. A few students felt disappointed with the competition as an evaluation methodology. One of our objectives was to qualify if competition is suitable from the evaluation point of view. We disregard the "0 to all" answers here because we find these to be biased responses, driven by frustration and lack of interest as opposed to the perception of the learning environment. When looking at the remaining Q2 answers we conclude that the project objectives were considered suitable by the majority of students. Overall the students found the challenge well organized. We consider these results encouraging when taking in consideration the limited number of students that have experienced the proposed methodology in two semesters.

### B. Evaluation Based on Statistical and Correlation Results

Validating the methodology can also be done through the statistical analysis of students' grades. This approach essentially checks the suitability of the grades the students obtained. We expect and assume a good accumulated correlation among the grades for the different class activities. We take benefit of the SRE evaluation process, which is organized around 5 activities: two theory exams, LAB1 evaluated using an individual lab exam, LAB2 evaluated based on the qualification round result and LAB3 evaluated based on the Champions League challenge, as described before. For our statistical analysis we correlate the grades across the two exams as the reference and across exams and labs. A good comparison would be between the average exam grade (EX_aver) and the LAB3 grade (LAB3-EX_aver). Such comparison assumes that EX_aver grade is a good measure of the learning process despite the differences in the learning principles between theory and lab.

We propose using the correlation coefficient R or the Pearson product moment correlation coefficient, which measures the strength and the direction of a linear relationship between two variables. We use it to compare the students grades obtained in two different evaluation sessions. Values for Pearson coefficient can go from -1 (strong downhill linear relationship) to 1 (strong uphill linear relationship), where -0.5 or 0.5 is considered as a moderate relationship and 0 as no linear relationship.

Before analyzing this class, we establish a baseline by analyzing a different subject of the same degree with a similar approach in terms of exams and labs. Those labs do not apply PBL or a competitive learning methodology. We analyze two different semesters, Sa and Sb. The student population is not the same but shares similar background and skills. Fig. 4 shows the scatter plots of student grades obtained through one evaluation item vs. another, where itemA-itemB indicates itemA on the y-axis and itemB on the x-axis. Observe the differences in the dispersion for the same itemA-itemB and the same subject taught by the same instructor in different semesters. Notice also the differences in the linear regression for each cloud of points, which seems to show the differences related to the peculiarities of the group of students.

If we look at the LAB3-EX_aver trend lines we observe that the slope is similar in both semesters, 0.7 and 1.3, whereas the offset is around 2 and -0.28, respectively. A slope below 1.0 indicates that LAB3 gives lower grades than the exams, assuming an offset close to 0. Table IV indicates the R coefficient for the previous set of values, where for most grades we observe a moderate to low linear positive relationship. Observe the similar value obtained for the R coefficient when comparing the averaged exam mark with the LAB3 mark: 0.550 and 0.785. This indicates that the grading obtained from LAB3 is comparable (moderate to high) to the average exams grade. If we consider the average exam grade as a representative evaluation parameter, LAB3 would be representative as well.

Fig. 5 shows the equivalent scatter points between labs and exams for the SRE subject that implements the Champions League methodology. When comparing the cloud of scatter points of Fig.5 with those of the reference subject of Fig. 4, we observe a relative resemblance. In other words, the distribution of the subject of interest aligns well with the results obtained from the reference subject. The LAB3-EX_aver trend line has a slope of 1.2 and an offset of 1.77. This is similar to the values obtained in the reference plots. The Champions slope is slightly higher than the performance (PER) slope, where both contribute to the LAB3 grades. Although it is difficult to identify which portion comes from the contribution of cooperative and competitive learning in LAB3 grades, we observe that this slightly higher slope motivates students to tackle the proposed activity.

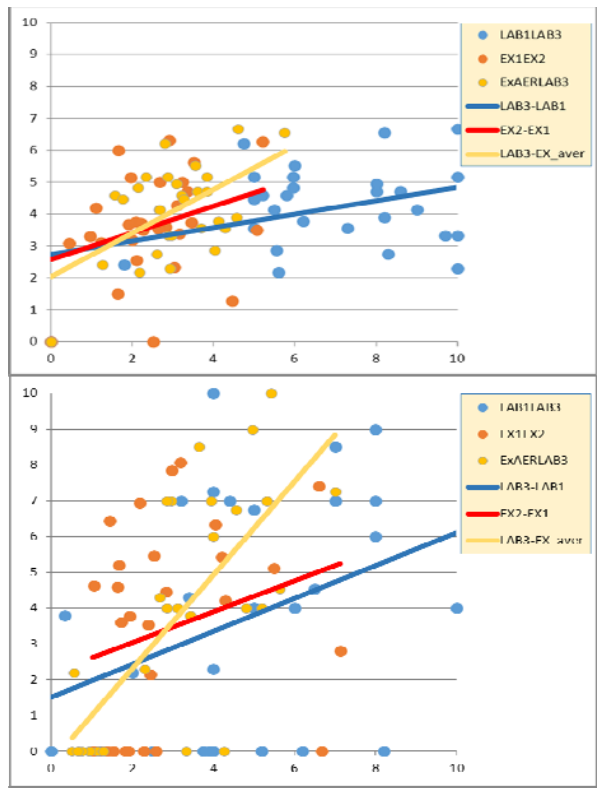

Fig. 4. Scatter and linear regression of evaluation item A (y-axis) over item B (x-axis). Course Sa is evaluated in the upper and Sb in the lower plot.

Table V indicates that correlation coefficient R exhibits similar values as those obtained for the reference subject in Table IV. These indicate that the overall process is coherent enough and that the proposed cooperative-competitive methodology shows at least similar learning capacities than traditional learning processes. Nevertheless, notice that the Champions coefficient R is 0.688, showing a higher correlation with the average of exams grades than the performance coefficient of 0.551. This suggests that the competitive part of the proposed learning methodology has a positive outcome on the grades.

Table IV. Pearson coefficients for reference subject.

| Course A | Ex2 | Lab3 |
|---|---|---|
| *Ex1* | 0.357 | 0.328 |
| *Ex2* | | 0.550 |
| *Exams Averaged* | | **0.550** |
| *Lab 1* | | 0.336 |
| **Course B** | **Ex2** | **Lab3** |
| *Ex1* | 0.345 | 0.370 |
| *Ex2* | | 0.822 |
| *Exams Averaged* | | **0.785** |
| *Lab 1* | | 0.382 |

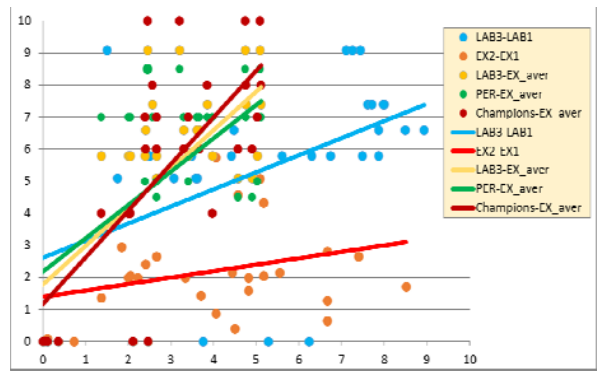

Fig. 5. Scatter and linear regression plots for the subject of interest with the proposed intervention.

### C. Grading Method Impact

Evaluating the impact of the grading method is another important aspect to keep in mind. The main objective is to analyze if the competitive learning approach appears to be considered as our main concern is the feasibility of the competition itself. Going further with this idea, the impact parameter I is calculated. This parameter is defined as the subtraction of the LAB3 grades and the weighted laboratory grades; where LAB3 is composed only of the Champions League marks and the weighted laboratory grades by the average of the marks of LAB1 and LAB3. This parameter indicates if LAB3 is positively or negatively affecting the student grades. Specifically, an impact parameter below zero means that the Champions grade are non-beneficial, whereas an impact parameter above zero suggests that LAB3 is beneficial for the students.

Fig. 6 plots the impact parameter I evaluated using a histogram. Specifically, histogram data (blue bars) fit a normal density function (red trace) with a mean of 0.02 and a standard deviation of 0.498. As can be seen, students are located mostly at the center of the histogram and the distribution reassembles a Gaussian profile centered around zero. This implies that, in general, the Champions method neither helps nor harms the students. Note that due to the limitations on the number of assets evaluated, the histogram shows some unfilled zones.

Furthermore, it is interesting to study the limit cases. Even if in general the method seems appropriate, it could be that it helps or harms specific cases. To do that, we look at the blue scattering points of Fig. 5, where the x-axis represents LAB1 grades and y-axis LAB3 grades. We observe that most points are in the mid to mid-right corner of the graph, meaning that the LAB1 and LAB3 grades are similar. However, there are three students whose LAB1 marks are fine, but 0 for LAB3. These corner cases are because these students dropped out after completing LAB1. There are still a few additional students whose LAB3 performance was lower than that of LAB1 (upper-right area of the figure). We identify two possible explanations: First, students who obtain very good grades in LAB1 do not bother to receive excellent grades in LAB3.

They are less motivated than others who need to improve (law of the minimum effort). The second explanation is related to the evaluation method. In the presence of competition, the emotional intensity increases, reinforcing cooperation and fierce competition at the same time. Nevertheless, we considered this as a minor negative impact.

On the other hand, there are a few students for whom the challenge was very beneficial (blue scattering points upper-left zone of Fig. 5), giving them the chance to turn things around and improve their lab grades. Overall, as mentioned in section VI-B, the blue scattering points of Fig. 5 depict a moderate-low correlation between LAB1 and LAB3 marks, even though the averages of 5.38 and 5.48 for LAB1 and LAB3, respectively, are very similar. Obtaining a good result in the Champions League challenge is enabled by the grading system and giving the evaluation method legitimacy, since the competition does not have fixed grades for fixed exercises as opposed to what is the case for traditional, non-competitive laboratory sessions. Moreover, as indicated by Fig. 6, this method does not provide grade improvement for free, which would be the case if the fitted curve were centered at a positive value. Neither is the method too strict, which would be the case if it were centered at a negative value.

Finally, the method itself provides self-motivation and offers a range of possibilities and encourages the students to carry on with the course because it gives them the opportunity to recover from slow starts.

## VII. DISCUSSION AND CONCLUSIONS

We proposed a cooperative-competitive learning intervention for teaching wireless communications and analyzed the proposed PBL methodology from different viewpoints. Despite the limited samples, we noticed that some students are fully against competition as a mechanism for education and grading. On the other hand, the feedback we received for the project objectives and the organization are encouraging. And, through our interaction with students, we noticed that some of them gained a valuable experience beyond the learning objectives. In the next editions of the SRE Champions League challenge, we will emphasize on the benefits of competitive learning and refine the methodology to improve motivation and satisfaction. One approach is to add mid-semester surveys to be able to address the needs of students early on.

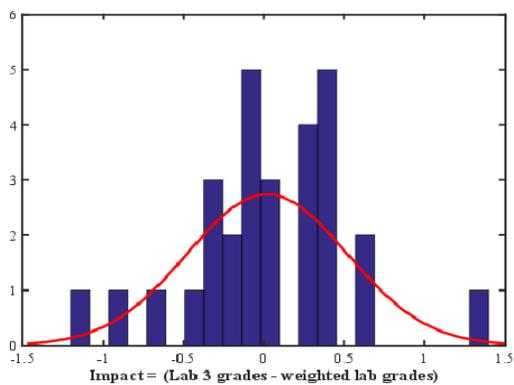

Fig. 6. Histogram of the impact parameter.

Table V. Pearson coefficients for SRE.

| SRE Course | Ex2 | Lab3 | Perform. | Champ. |
|---|---|---|---|---|
| *Ex1* | 0.291 | 0.607 | | |
| *Ex2* | | 0.377 | | |
| *Exams Aver.* | | 0.637 | 0.551 | 0.688 |
| *Lab 1* | | 0.493 | | |

One of the main conclusions that we draw is the need to improve the SDR tools and associated documentation. The learning curve of SDR development tools that allow real-time processing is too steep to be covered in a single course, unless the tools are truly user-friendly, the objectives well confined and the students well prepared. This class looks at the complex interactions between radio and computing resources. Being an experimental real-time system, computing resources are limited and, consequently, both the algorithmic correctness and implementation efficiency need to be considered by design and not as an afterthought. The students need to check system details not only related to the radio communications part, but also those that are associated with the real-time execution on practical platforms, as opposed to super-computers or non-real time simulators. For the next editions we plan to incorporate these lessons to improve the ALOE tools.

Regarding the suitability of the grading we found a good correlation with other teaching and evaluation methodologies. More precisely, we have shown that correlation results are in consonance with grading correlations between different items in the same course and across courses. This suggests that the proposed cooperative-competitive learning methodology promises coherent learning outcomes. In addition, we analyzed the impact of the proposed learning and grading methodology on grading fairness. Our initial analysis has shown that the methodology neither helps nor harms the students. More data needs to be acquired to make a definite conclusion.

We also observed certain benefits that we have not originally envisaged. One is the strong collaboration among different teams. We observed this early on and until a couple of weeks before the playoffs. Then, collaboration turned into competition and little information has been shared among the teams. We find this being a natural evolution facilitated by the intervention. After a cooperative learning phase, teams get busy learning about the system details, options and peculiarities of real SDR systems while developing their team strategies for the playoffs.

We conclude that the proposed PBL cooperative-competitive methodology using the Champions League format a suitable for teaching wireless communications to engineering students, but it requires careful preparation and continuous adaptation to address the student needs and technology evolution.

## ACKNOWLEDGMENT

The work done at Barcelona Tech was supported by the Spanish Government (MINECO) and FEDER under project TEC2017-83343-C4-2-R.

## REFERENCES

[1] R. M. Felder, G. N. Felder, and E. J. Dietz, "A longitudinal study of engineering student performance and retention. V. comparisons with traditionally-taught students," J. Eng. Educ., vol. 87, no. 4, pp. 469–480, Oct. 1998.

[2] B. Timmerman and R. Lingard, "Assessment of active learning with upper division computer science students," in Proc. 33rd ASEE/IEEE Frontiers in Educ. Conf., 2003, pp. S1D–7.

[3] M. J. Prince and R. M. Felder, "Inductive teaching and learning methods: Definitions, comparisons, and research bases," J. Eng. Educ., vol. 95, no. 2, pp. 123–138, 2006.

[4] R. Pucher, A. Mense, and H. Wahl, "How to motivate students in project based learning," in Proc. 6th IEEE AFRICON, vol. 1. George, South Africa, Oct. 2002, pp. 443–446.

[5] C. L. Dym, A. M. Agogino, O. Eris, D. D. Frey, and L. J. Leifer, "Engineering design thinking, teaching, and learning," J. Eng. Educ., vol. 94, no. 1, pp. 103–120, 2005.

[6] S. Boss and J. Krauss, "Reinventing Project-Based Learning. Your Field Guide to Real-World Projects in the Digital Age," Eugene, OR, USA: Int. Soc. Technol. Educ., 2007.

[7] A. Gelonch, V. Marojevic, I. Gomez, "Teaching telecommunications standards: bridging the gap between theory and practice," IEEE Commun. Mag., Vol. 55, Iss. 5, pp. 145-153, May 2017.

[8] D. W. Johnson and R. T. Johnson, "Learning together and alone: cooperative, competitive and individualistic learning." 5th edition. Needham Heights, MA: Allyn & Bacon, 1998.

[9] D. W. Johnson, & R. T. Johnson, "Cooperative learning and classroom and school climate," Educational environments: Evaluation, antecedents and consequences, pp. 55-74, 1991.

[10] D. T. Rover, R. A. Mercado, Z. Zhang, M. C. Shelley and D. S. Helvick, "Reflections on teaching and learning in an advanced undergraduate course in embedded systems," IEEE Trans. Educ., vol. 51, no. 3, pp. 400-412, Aug. 2008. doi: 10.1109/TE.2008.921792.

[11] L. Jing, Z. Cheng, J. Wang and Y. Zhou, "A spiral step-by-step educational method for cultivating competent embedded system engineers to meet industry demands," IEEE Trans. Educ., vol. 54, no. 3, pp. 356-365, Aug. 2011. doi: 10.1109/TE.2010.2058576.

[12] D. J. Cappelleri and N. Vitoroulis, "The robotic decathlon: Project-based learning labs and curriculum design for an introductory robotics course," IEEE Trans. Educ., vol. 56, no. 1, pp. 73–81, Feb. 2013.

[13] R. W. Y. Habash and C. Suurtamm, "Engaging high school and engineering students: A multifaceted outreach program based on a mechatronics platform," IEEE Trans. Educ., vol. 53, no. 1, pp. 136–143, Feb. 2010.

[14] M. C. Rodriguez-Sanchez, A. Torrado-Carvajal, J. Vaquero, S. Borromeo, and J. A. Hernandez-Tamames, "An embedded systems course for engineering students using open-source platforms in wireless scenarios," IEEE Trans. Educ., vol. 59, no. 4, pp. 248–254, Nov. 2016.

[15] O. Arbelaitz, J. I. Martin, and J. Muguerza, "Analysis of introducing active learning methodologies in a basic computer architecture course," IEEE Trans. Educ., vol. 58, no. 2, pp. 110–116, May 2015.

[16] H. Hassan et al., "A multidisciplinary PBL robot control project in automation and electronic engineering," IEEE Trans. Educ., vol. 58, no. 3, pp. 167–172, Aug. 2015.

[17] I. Calvo, I. Cabanes, J. Quesada and O. Barambones, "A Multidisciplinary PBL Approach for Teaching Industrial Informatics and Robotics in Engineering," in IEEE Transactions on Education, vol. 61, no. 1, pp. 21-28, Feb. 2018.

[18] L. Smith and S. Mann, "Playing the game: A model for gameness in interactive game based learning," in Proc. 15th Annual Nat. Advisory Committee on Computing Qualifications, Hamilton, New Zealand, 2002, pp. 397–402.

[19] K. Becker, "Teaching with games: The minesweeper and asteroids experience," J. Comput. Small Colleges, vol. 17, no. 2, pp. 23–33, Dec. 2001.

[20] A. Kohn, No Contest: The Case Against Competition. Boston, MA, USA: Houghton Mifflin, 1986.

[21] R. Lawrence, "Teaching data structures using competitive games," IEEE Trans. Educ., vol. 47, pp. 459–466, Nov. 2004.

[22] H. J. Brightman, GSU Master Teacher Program: On Critical Thinking Georgia State Univ., 2006 [Online]. Available: http://www2.gsu.edu/~dschjb/wwwcrit.html [Accessed 5/2/2018].

[23] F. Y. Yu, "Reflections upon cooperation-competition instructional strategy: Theoretical foundations and empirical evidence," Nat. Chi Nan Univ. J., vol. 5, no. 1, pp. 181–196, 2001.

[24] Z. H. Chen, C. Liao, and T. W. Chan, "Learning by pet-training competition: Alleviating negative influences of direction competition by training pets to compete in game-based environments," in Proc. IEEE Int. Conf. Adv. Learning Technol., 2010, pp. 411–413.

[25] L. J. Chang, J. C. Yang, F. Y. Yu, and T. W. Chan, "Development and evaluation of multiple competitive activities in a synchronous quiz game system," J. Innovat. Educ. Training Int., vol. 40, no. 1, pp. 16–26, Jan. 2003.

[26] A. Collins, J. S. Brown, and S. E. Newman, Cognitive Apprenticeship: Teaching the Crafts of Reading, Writing, and Mathematics, Knowing, Learning, and Instruction: Essays in Honor of Robert Glaser. Hillsdale, NJ, USA: Erlbaum, 1989, pp. 453–494.

[27] J. D. Brown and M. A. Marshall, "Self-esteem and emotion: Some thoughts about feelings," Pers. Social Psychol. Bull., vol. 27, no. 5, pp. 575–584, May 2001.

[28] D. A. Stapel and W. Koomen, "Competition, cooperation, and the effects of others on me"," J. Pers. Social Psychol., vol. 88, pp. 1029–1038, 2005.

[29] A. Siddiqui, M. Khan, and S. Katar, "Supply chain simulator: A scenario-based educational tool to enhance student learning," Comput. Educ., vol. 51, no. 1, pp. 252–261, Aug. 2008.

[30] M. Fasli and M. Michalakopoulos, "Supporting active learning through game-like exercises," in Proc. 5th IEEE Int. Conf. Adv. Learn. Technol., 2005, pp. 730–734.

[31] T. A. Philpot, R. H. Hall, N. Hubing, and R. E. Flori, "Using games to teach statics calculation procedures: Application and assessment," Comput. Appl. Eng. Educ., vol. 13, no. 3, pp. 222–232, 2005.

[32] F. Y. Yu, L. J. Chang, Y. H. Liu, and T. W. Chan, "Learning preferences towards computerized competitive modes," J. Comput. Assist. Learn., vol. 18, no. 3, pp. 341–350, Sep. 2002.

[33] J. Dettmer. "Competition photography... learning by losing," PSA Journal, 36, June 2005.

[34] C. H. Lin, S. H. Huang, J. L. Shih, A. Covaci and G. Ghinea, "Game-based learning effectiveness and motivation study between competitive and cooperative modes," in Proc. 2017 IEEE 17th International Conference on Advanced Learning Technologies (ICALT), Timisoara, 2017, pp. 123-127. doi: 10.1109/ICALT.2017.34

[35] B. Silva and R. N. Madeira, "A study and a proposal of a collaborative and competitive learning methodology," in Proc. IEEE Eng. Educ. Conf., 2010, pp. 1011–1018.

[36] I. Gomez, V. Marojevic, A. Gelonch, "ALOE: an open-source SDR execution environment with cognitive computing resource management capabilities," IEEE Commun. Mag., Vol. 49, Iss. 9, pp. 76-83, Sept. 2011.

[37] http://www.crlt.umich.edu/resources/evaluation-teaching/methods [Accessed 5/2/2018]